\documentstyle[preprint,aps,floats,epsfig]{revtex}

\tightenlines
\begin{document}

\preprint{\vbox{\hbox{JLAB-THY-01-02}}}

\title{Quark-Hadron Duality in Structure Functions}

\author{ Nathan Isgur$^{(1)}$, Sabine Jeschonnek$^{(1)}$,
    W. Melnitchouk$^{(1,2)}$, and J. W. Van Orden$^{(1,3)}$}

\address{$^{(1)}$   Jefferson Lab,
            12000 Jefferson Ave,
            Newport News, VA 23606}
\address{$^{(2)}$   Special Research Centre for the
            Subatomic Structure of Matter,  \\
            Adelaide University,
            Adelaide 5005, Australia}
\address{$^{(3)}$   Department of Physics,
            Old Dominion University,
            Norfolk, VA 23529}

\maketitle

\begin{abstract}

While quark-hadron duality is well-established experimentally, the
current theoretical understanding of this important phenomenon is
quite limited.  To expose the essential features of the dynamics
behind duality, we use a simple model in which the hadronic spectrum
is dominated by narrow resonances made of valence quarks.  We
qualitatively reproduce the features of duality as seen in electron
scattering data within our model. We show that in order to observe
duality, it is essential to use the appropriate scaling variable and
scaling function.  In addition to its great intrinsic interest in
connecting the quark-gluon and hadronic pictures, an understanding of
quark-hadron duality could lead to important benefits in extending the
applicability of scaling into previously inaccessible regions.

\end{abstract}


\newpage

\section{Introduction}

\subsection{Background}

Duality is a much used and much abused concept.
In some cases it is used to describe an equivalence between quark- and
hadron-based pictures which is trivial; in others an equivalence which
is impossible.
In almost all cases, the conceptual framework in which duality is
discussed and used is either hopelessly muddled or hopelessly abstract.
Nevertheless, the data indicate that some extremely interesting and
potentially very important ``duality'' phenomena are occurring at low
energy.

     We begin by making the trivial observation that any hadronic
process can be correctly described in terms of quarks and gluons,
assuming that Quantum Chromodynamics (QCD) is the correct theory for
strong interactions.  While this statement is obvious, it rarely has
practical value, since in most cases we can neither perform nor
interpret a full QCD calculation.  We will refer to the above
statement that any hadronic process can be described by a full QCD
calculation as ``degrees of freedom duality'': if one could perform
and interpret the calculations, it would not matter at all which set
of states --- hadronic states or quark and gluon states --- was used.

    On the other hand, there are rare cases where the average of
hadronic observables is described by a perturbative QCD (pQCD)
calculation. We reserve the use of the term ``duality'' to describe
these rare correspondences, in contrast to the trivial ``degrees of
freedom duality'' described above. In these rare cases, a quark-gluon
calculation leads to a very simple description of some phenomenon even
though this phenomenon ``materializes'' in the form of hadrons. Deep
inelastic scattering is the prototypical example, and the one on which
we focus here. These rare examples are all characterized by a special
choice of kinematic conditions which serve to expose the ``bare''
quarks and gluons of the QCD Lagrangian.  In the case of deep
inelastic scattering, the kinematics are such that the struck quark
receives so much energy over such a small space-time region that it
behaves like a free particle during the essential part of its
interaction. This leads to the compellingly simple picture that the
electron-nucleon cross section is determined in this kinematic region
by free electron-quark scattering, i.e. duality is exact for this
process in this kinematic regime.

      For inclusive inelastic electron scattering from a proton in the
scaling region, the cross section is determined by the convolution
of a non-perturbative and currently difficult to calculate parton
distribution function with an electron-quark scattering cross
section determined by perturbative QCD (pQCD). For semileptonic
decays of heavy quarks, {\it e.g.}  $\bar{B} \rightarrow X_c l
\bar \nu_l$, one can prove using pQCD that the decay rate is
determined by that of the underlying heavy quark, in this case
obtained from the process $b \rightarrow c l \bar \nu_l$
\cite{isgurwise}.  In $e^+e^- \rightarrow hadrons$, it is the
underlying $e^+e^- \rightarrow q \bar q$ process that applies
because of pQCD.  However, while duality applies to all of these
phenomena, we will see that even in these special processes we
must invoke an averaging procedure to identify the hadronic
results with the quark-gluon predictions.

     In addition to its need of an averaging procedure, it is easy to
see that the pQCD picture of inelastic electron scattering must fail
for $Q^2 \rightarrow 0$. For duality to hold for the nucleon structure
functions in this case, the elastic electric proton and neutron form
factors, which take the value of the nucleon charges for $Q^2
\rightarrow 0$, would have to be reproduced by electron scattering off
the corresponding $u$ and $d$ quarks.  This is possible for the proton
since the squares of the charges of two $u$ quarks and one $d$ quark
add up to 1 \cite{gottfried}. However, for the neutron, the squared
quark charges cannot add up to 0, so it is clear that local duality in
inclusive inelastic electron scattering from a neutron must fail for
$Q^2 \rightarrow 0$.  Also, we know that duality must fail for
polarized structure functions at low $Q^2$, as the Ellis-Jaffe sum
rule and the Gerassimov-Drell-Hearn sum rule, which can be written as
integrals over $g_1(\nu,Q^2)$ at different $Q^2$, are negative (GDH
sum rule for $Q^2 = 0$) and positive (Ellis-Jaffe sum rule at $Q^2$ of
several GeV), respectively \cite{g1dual}.

     Thus duality in   inelastic electron scattering has to hold in the
scaling regime and must in general  break down at low energy.
Obviously, a very interesting question is what happens in between these
regimes, {\it i.e.} how does duality break down?
This paper answers this question, which  is not only interesting in itself,
but also crucial for  practical, quantitative applications of
duality.

\subsection{Introducing Local Averaging and Our Model}

   We begin by discussing   the issue of averaging.  If duality is relevant
at all at low energy, then  it is quite obvious that we need to perform some
sort of average: the smooth, analytic pQCD
prediction cannot in general correspond exactly to the generally highly
structured hadronic data.  For low energies this requirement is universally
accepted; however, even in the ``scaling'' region one must
average in principle. To see this, consider QCD in the large-$N_c$ limit
\cite{largenc}. We can do this because no element of the pQCD results for
deep inelastic scattering depends on the number of colors. However, in this
limit the hadronic spectrum consists entirely of infinitely narrow
noninteracting resonances
\cite{baryoncaveat}, {\it i.e.},
there are only infinitely narrow spikes in the
$N_c \rightarrow \infty$ hadronic world.
Since the quark level calculation still yields a smooth scaling curve, and the kinematic conditions for being in the scaling region  are unchanged as
$N_c \rightarrow \infty$,  we see that we must average even in
the scaling region. While in Nature, the resonances have fairly broad decay
widths so that  the averaging takes place automatically in the data, the
large
$N_c$ limit shows us that averaging is always required in principle. It is
thus clearly important to be able to define this averaging procedure, {\it
{\it e.g.}}, how large the intervals must be  and which resonances have to
be included.

      It is easy to see that this procedure will not be universal, and
will certainly not simply be that the resonances one-by-one locally
average the pQCD-derived scaling curve: the averaging method will
depend on the process and on the target. Consider, as an illustration
of these points, the case of a spinless quark and antiquark with
charges $e_1$ and $e_2$ and equal masses bound into a nonrelativistic
$q_1 \bar q_2$ system.  The inelastic electron scattering rate
calculated at the quark level in leading twist will then be proportional to
$e^2_1+e^2_2$. Since the elastic state will be produced with a rate
proportional to $(e_1+e_2)^2$, it clearly cannot in general be locally
dual to the scaling curve \cite{schmidt}. How then is duality realized in this
system? Consider the charge operator $\sum_i e_i e^{i\vec q \cdot \vec
r_i}$: from the ground state it excites even partial wave states with
an amplitude proportional to $e_1+e_2$ and odd ones with an amplitude
proportional to $e_1-e_2$. Thus the resonances build up a cross
section of the form $\alpha_1 (e_1+e_2)^2 + \alpha_2 (e_1-e_2)^2 +
\alpha_3 (e_1+e_2)^2 + \cdots$ and one can see by explicit calculations
in models that (up to phase space factors) the cross terms in this sum
will cancel to give a cross section proportional to $e^2_1+e^2_2$ once
averaged over nearby even and odd parity resonances. It is clear that
such target- and process-dependence is worthy of study.  However, in
this paper we will restrict ourselves to a model with $e_2=0$ so that
local duality might apply \cite{CloseIsgur}.

   The question of the validity  of low energy duality, {\it i.e.}, duality
in electron scattering at finite beam energies in  inelastic
electron scattering after suitable averaging, is as old as the
first inclusive electron scattering experiments themselves. It
begins with the seminal paper of Bloom and Gilman
\cite{bgduality}, which made the observation that the inclusive
$F_2$ structure function in the  resonance region at low $Q^2$
generally oscillates about and averages to a global scaling curve
which describes high $Q^2$ data. More recently, interest in
Bloom-Gilman duality has been revived with the collection of high
precision data on the $F_2$ structure function from Jefferson Lab
\cite{JLAB}. These data not only confirmed the existence of
Bloom-Gilman duality to rather low values of $Q^2$, but also seem
to demonstrate  that for the proton the equivalence of the
averaged resonance and scaling structure functions holds also for
each  resonance so that     duality also exists locally.

Here we present a model for the study of quark-hadron duality in
electron scattering that uses only a few basic ingredients. Namely, in
addition to requiring that our model be relativistic, we assume
confinement and that it is sufficient to consider only valence quarks
(this latter simplification being underwritten, as mentioned
previously, by the large $N_c$ limit).  In addition, since our model
is designed to explore conceptual issues and not to be compared to
data, and since we postpone addressing spin-dependent issues to later
work, for simplicity we also take the quarks, electrons and photons to
be scalars. A model with these features will not give a realistic
description of any data, but it should allow us to study the critical
questions of when and why duality holds.  While this model is
extremely simple, we see no impediment to extending it to describe a
more realistic situation since we find that duality arises from the
most basic properties of our model.

   We make several more convenient simplifications. Although it is our aim
to study duality in electron scattering from the nucleon, {\it
i.e.} from a three-quark-system, as a first step we study these
issues in what is effectively a one quark system by considering
such a quark to be confined to an infinitely massive antiquark. In
the case of scalar quarks considered here, we can therefore
describe the system by the Klein-Gordon equation. We also select
for our  confining potential one which is linear in $r$, namely
$V^2 (\vec r \,) = \alpha \, r^2$, where $\alpha$ is a
generalized, relativistic string constant. This choice allows us
to obtain analytic solutions, without which the required numerical
work for this study would be daunting. Indeed, the energy
eigenvalues, $E_N = \sqrt{ 2 \sqrt{\alpha}(N + 3/2) + m^2 }$,
where $m$ is the mass of interacting quark, can be readily
obtained by noting the similarity to the Schr\"odinger equation
for a non-relativistic harmonic oscillator potential: the
solutions for the wave functions are the same as for the
non-relativistic case.

    In the next Section we construct the structure function out of
resonances described by form factors, each of which individually gives
vanishing contributions at large momenta, and show that it both scales
and, when suitably averaged, is equal to the ``free quark'' result.
An analysis in terms of structure function moments is presented in
Section III. In Section IV we examine the onset of scaling, and the
appearance of Bloom-Gilman duality, while in Section V we discuss the
connection of Bloom-Gilman duality with duality in heavy quark
systems.  Finally, in Section VI we summarize our results and mention
some possible directions for future research.

\section{Quark-Hadron Duality in the Scaling Limit}

The differential cross section for inclusive inelastic scattering of a
``scalar
electron'' {\it via} the exchange of a ``scalar photon'' is
\begin{equation}
\frac{d \sigma} {dE_f d\Omega_f} = \frac{g^4}{16\pi^2}
\frac{E_f}{ E_i} \frac{1}{Q^4}\ {\cal W} \,,
\label{defscalw}
\end{equation}
where the scalar coupling constant $g$ carries the dimension of a
mass, and the factor multiplying the scalar structure function ${\cal
W}$ corresponds to the Mott cross section.  In a model where the only
excited states are infinitely narrow resonances, ${\cal W}$ is given
entirely by a sum of squares of transition form factors weighted by
appropriate kinematic factors:
\begin{equation}
{\cal W}(\nu,\vec q \, ^2)
= \sum_{N=0}^{N_{\rm max}} \, \frac{1}{4 E_0 E_N}\, \,
\left| F_{0N}(\vec q) \right|^2 \, \, \delta(E_N - E_0 - \nu)\, ,
\label{wscalho}
\end{equation}
where $\vec q \equiv \vec p_i - \vec p_f$, the form factor $F_{0N}$
represents a transition from the ground state to a state characterized
by the principal quantum number $N$, and the sum over states $N$ goes
up to the maximum $N_{\rm max}$ allowed kinematically. Note that for
fixed, positive $Q^2 \equiv \vec q \, ^2 - \nu^2$, $N_{max} = \infty$.

    The excitation form factors can be
derived using the recurrence relations of the Hermite polynomials.
One finds:
\begin{equation}
\label{eqff}
F_{0N} (\vec q \, ^2) = \frac{1}{\sqrt{N!}} \, i^N \,
\left ( \frac{|\vec q|}{\sqrt{2} \, \beta} \right ) ^N
\exp (- \vec q \, ^2 / 4 \, \beta^2) \, ,
\end{equation}
where $\beta = \alpha^{1/4}$. This form factor is in fact the sum
of all form factors for excitations from the ground state to
degenerate states with the same principal quantum number $N$. As a
precursor to our discussion of duality, we note that it will be a
necessary condition for duality that these form factors (or more
generally those corresponding to some other model potential) can
represent the pointlike free quark. It is in fact the case that
$\sum_{N=0}^{N_{\rm max}} |F_{0N} (\vec q \,)|^2 \rightarrow 1$ as
$N_{\rm max} \rightarrow \infty$, a relation which follows from
the completeness of the confined wave functions. Incidentally, an
examination of the convergence of this sum as a function of $\vert
\vec q \vert^2$ is sufficient to make the point that reproducing
the behavior of a free quark requires more and more resonances as
$\vert \vec q \vert^2$ increases (details of this will be
discussed in a forthcoming publication).

   Scaling in the presence of confining final
state interactions has previously been investigated in Refs.
\cite{ioffe,gurvitzrinat,greenberg,psl}, where similar conclusions
are reached. This suggests that scaling may indeed be a trivial
feature of a large class of simple quantum mechanical models. Some
sense of how this can occur can be obtained by considering some of
the properties of the relativistic oscillator model used in this
paper. In particular, consider the properties of the square of the
form factors.  For a fixed principal quantum number, $N$, the form
factor has a maximum in $|\vec q|$ at $\vec q \,
^2_N=2\beta^2\,N$. Using $\nu_N=E_N-E_0$ and
$E_N=\sqrt{2\beta^2\,N+E_0^2}$, it can be shown that
\begin{equation}
\nu_N=\frac{Q^2_N}{2E_0}
\end{equation}
where $Q^2_N=\vec q \, ^2_N - \nu_N^2$. So the position of the peak in
the averaged structure function occurs at $u_{Bj}={m/E_0}$ where
$u_{Bj}={Q^2/2m\nu}$ is a scaled Bjorken scaling variable
$u_{Bj}\equiv\frac{M}{m}x_{Bj}$ which takes into account that as the
mass of the antiquark $M_{\bar Q}\rightarrow\infty$, the constituent
quark will carry only a fraction of order $m/E_0$ of the hadron's
infinite-momentum-frame momentum. Furthermore, for fixed $\vec q$ the
structure function falls off smoothly for energy transfers away from
the peak value.  The width of this peak as a function of energy
transfer also becomes constant for large $|\vec q|$.

Now consider the integral of the structure function
\begin{equation}
\Sigma(\vec q \, ^2)=\int_0^\infty d\nu\ {\cal W}(\nu,\vec q \,
^2) =\sum_{nlm}\frac{1}{4E_0E_N}<\psi_{000}|\rho(-\vec
q)|\psi_{nlm}> <\psi_{nlm}|\rho(\vec q)|\psi_{000}>
\end{equation}
where $N=2(n-1)+l$ with $n=1,2,3,\cdots$, and where $\rho(\vec
q)=e^{i\vec q\cdot\vec x}$. Since the form factor sum for a fixed
$\vec q$ peaks about $E_{N_{max}}=\sqrt{\vec q \, ^2+E_0^2}$, we
can substitute $E_N\rightarrow E_{N_{max}}$ and then sum over the
complete set of final states to give
\begin{equation}
\Sigma(\vec q \, ^2)\cong\frac{1}{4E_0E_{N_{max}}}\cong\frac{1}{4E_0q}
\end{equation}
for large momentum transfer. Therefore, if we define the scaling
function as ${\cal S}\equiv |\vec q|\ {\cal W}$, as will be done
below, the area of the scaling function becomes constant at large
momentum transfer.

Since the scaling function peaks at fixed $u_{Bj}$, smoothly falls
about the peak, has fixed width and constant area at large
momentum transfer, the model scales.  It is a common misconception
that the presence of scaling implies that the final states must
become plane waves.  In fact, the argument above makes it clear
that scaling occurs when the structure function becomes
independent of the final states as in the closure approximation
used here.

   To see duality clearly both experimentally and theoretically, one
needs to go beyond the Bjorken scaling variable $x_{Bj}$ and the
scaling function ${\cal{S}}_{Bj} = \nu \cal{W}$ that goes with it.
This is because in deriving Bjorken's variable and scaling
function, one not only assumes $Q^2$ to be larger than any mass
scale in the problem, but also that high $Q^2$ (pQCD) dynamics
controls the interactions. However, duality has its onset in the
region of low to moderate $Q^2$, and there masses and violations
of asymptotic freedom do play a role. Bloom and Gilman used a
new, {\it ad hoc} scaling variable $\omega'$ \cite{bgduality} in
an attempt to deal with this fact. In most contemporary data
analyses, the Nachtmann variable \cite{greenbergb,nachtmann} is
used together with ${\cal{S}}_{Bj}$. Nachtmann's variable contains
the target mass as a scale, but neglects quark masses. For our
model, the constituent quark mass (assumed to arise as a result of
spontaneous chiral symmetry breaking) is vital at low energy, and
a scaling variable that treats both target and quark masses is
desirable. Such a variable was derived more than twenty years ago
by Barbieri {\it et al.}  \cite{barbieri} to take into account the
masses of heavy quarks; we use it here given that after
spontaneous chiral symmetry breaking the nearly massless light
quarks have become massive constituent quarks, calling it
$x_{cq}$:
\begin{equation}
x_{cq} = \frac{1}{2 M} \left ( \sqrt{\nu^2 + Q^2} - \nu \right )
\left ( 1 + \sqrt{1 + \frac{4 m^2}{Q^2}} \right ) \, .
\label{defxdis}
\end{equation}
The scaling function associated with this variable is given by:
\begin{equation}
\label{S}
  {\cal{S}}_{cq} \equiv |\vec q|\ {\cal W} = \sqrt{\nu^2 + Q^2}\ {\cal W}\,.
\end{equation}
This scaling function and variable were derived for scalar quarks
which are free, but have a momentum distribution. The derivation
of a new scaling variable and function for bound quarks will be
published elsewhere. Numerically, this scaling variable does not
differ very much from the one in Eq. (\ref{defxdis}). Of course
all versions of the scaling variable must converge to $x_{Bj}$ and
all versions of the scaling function must converge towards
${\cal{S}}_{Bj}$ for large enough $Q^2$. One can also easily
verify that in the limit $m \to 0$ one obtains from
(\ref{defxdis}) the Nachtmann scaling variable. In the following,
we use the variable $x_{cq}$ and the scaling function
${\cal{S}}_{cq}$.

\begin{figure}[!h,t]    
\begin{center}
\leavevmode
\epsfig{file = 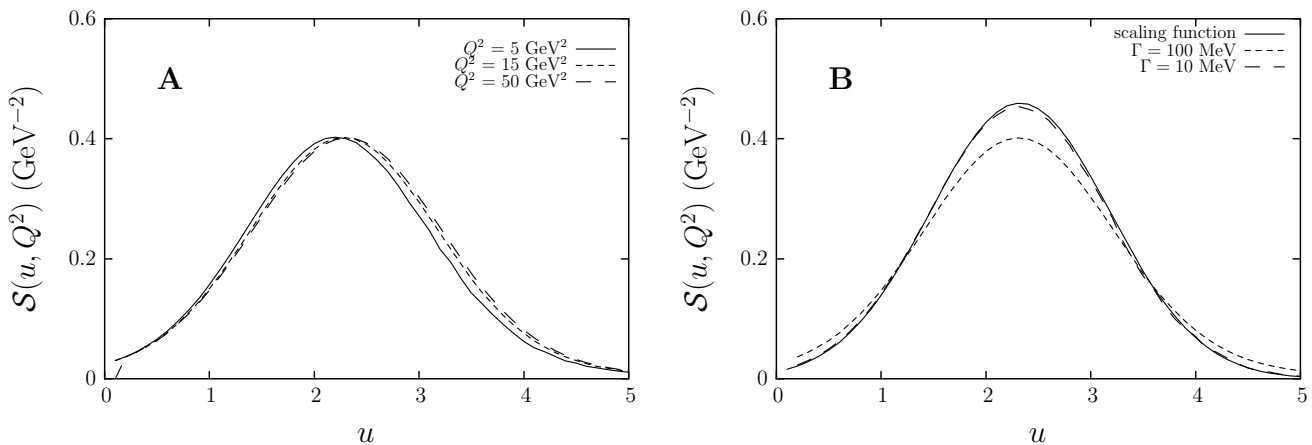, height=16cm}
\end{center}
\vspace{-10cm}
\caption{The high energy scaling behavior of ${\cal{S}}_{cq}$ as a
function of $u$ for various values of $Q^2$. In panel A we have used
$\Gamma = 100$~MeV to give the impression of real resonances even
though this large value distorts the scaling curve somewhat; for any
width equal to or smaller than this, the distortion is rather
innocuous, and for $\Gamma \rightarrow 0$, the structure function
approaches the scaling function in Eq.~(\ref{sfanalyt}), as shown in
panel B.}

\label{figscal1}
\end{figure}

   We are now ready to look at scaling and duality in our model. Since the
target has mass
$M
\rightarrow
\infty$, it is convenient to rescale the scaling variable $x_{cq}$   by a
factor
$M/m$:
\begin{eqnarray}
u & \equiv & {M \over m}\ x_{cq}\ ~~.
\end{eqnarray}
The variable $u$ takes values from 0 to a maximal, $Q^2$
dependent value, which can go to infinity.
The high energy scaling behavior of the appropriately rescaled structure
function
${\cal {S}}_{cq}$ is illustrated in Fig.~1.

The structure function has been evaluated using the
phenomenologically reasonable parameters $m = 0.33$~GeV and
$\alpha = (0.4 ~{\rm GeV})^{1/4}$, though we remind the reader not
to compare our results, which might resemble electron scattering
from a $B$ meson, to nucleon data! To display it in a visually
meaningful manner, the energy-dependent $\delta$-function has been
smoothed out by introducing an  unphysical Breit-Wigner shape with
an arbitrary but small width, $\Gamma$ chosen for purposes of
illustration:
\begin{equation}
\delta(E_N - E_0 - \nu) \rightarrow \frac{\Gamma}{2 \pi} \, \,
\frac{f}{(E_N - E_0 - \nu)^2 + (\Gamma/2)^2}\, ,
\end{equation}
where the factor
$
f = {\pi}/[{\frac{\pi}{2}
                + \arctan {2 (E_N - E_0) \over \Gamma}] }
$
ensures that the integral over the $\delta$-function is identical
to that over the Breit-Wigner shape.
The curves in Fig.~1  show that
scaling sets in rather rapidly.
The resonances show up as bumpy structures in the low $Q^2$ region
(which will be discussed in Section IV below), a trace of which is
visible for the $Q^2 = 5$~GeV$^2$ curve.

By  taking the continuum limit for
the energy and applying Stirling's formula, one can obtain an analytic
expression for the scaling curve, valid in the scaling region, for the
transition of the quark from the ground state to the sum of all excited
states:
\begin{equation}
{\cal{S}}_{cq}(u) = \frac{E_0}{  \sqrt{\pi} \beta }
        \exp{\frac{(E_0-m u)^2}{\beta^2}} \, .
\label{sfanalyt}
\end{equation}
Of course we still need to verify that this scaling curve as seen in
Fig.  1 found by summing over hadrons is the same as the one which we
would obtain from deep inelastic scattering off the quark, {\it {\it
i.e.}}, if we were to switch off the potential in the final state. In
this case, the tower of hadronic states is replaced by the free quark
continuum. Duality predicts that the results should be the same in the
scaling limit, and by direct calculation we confirm this.

\section{Moments of Structure Functions}
\label{secglo}

Bloom-Gilman duality relates structure functions at low and high $Q^2$
averaged over appropriate intervals of the hadronic mass $W$. As a
quantitative measure of this feature of the data, one  conventionally
examines the $Q^2$-dependence of moments of structure functions.
The moments offer the cleanest connection with the
operator product expansion of QCD, and provide a natural connection
between duality in the high- and low-$Q^2$ regions. By considering the
moments, we also remove   artifacts introduced through the smoothing
procedure described above for the structure function itself.

The moments of the structure function ${\cal{S}}_{cq}(u,Q^2)$ are defined as:
\begin{equation}
\label{moment}
M_n(Q^2) = \int_0^{u_{\rm max}} du \, \, u^{n-2} \,
        {\cal{S}}_{cq}(u,Q^2)\, ,
\end{equation}
where $u_{\rm max}$ corresponds to the maximum value of $u$ which is
kinematically accessible at a given $Q^2$.
 Evaluating the moments of the structure function (\ref{S})
explicitly  one has (provided the kinematics allow us to access all excited
states):
\begin{eqnarray}
M_n(Q^2) &=&
\left ( \frac{r}{2 m} \right )^{n-1} \,
\sum_{N=0}^{\infty} \,
\left( \sqrt{\nu_N^2 + Q^2} - \nu_N \right)^{n-1}
\, \, \frac{E_0}{ E_N} \,
\left|F_{0N} \left(\sqrt{\nu_N^2 + Q^2} \right) \right|^2 ~~ ,
\end{eqnarray}
where $\nu_N = E_N - E_0$ and $r = 1 + \sqrt{1 + 4 m^2/Q^2}$.
The elastic contribution to the moments is
\begin{equation}
M_n^{\rm elastic} (Q^2) = \left ( \frac{r}{2 m} \right )^{n-1}\,
Q^{n-1} \,  \, \left| F_{00} (Q^2) \right|^2\
= u^{n-1}_0 \left| F_{00} (Q^2) \right|^2\, ~~,
\end{equation}
where $u_0(Q^2)$ is the position in $u$ of the ground state. Note that
$M_n^{\rm elastic} (Q^2)$ becomes independent of
$n$ in the limit
$Q^2
\rightarrow 0$,
approaching unity and that
the inelastic contributions to the moments vanish for vanishing $Q^2$.

In Fig.~2 we show the $n=2$, 4, 6 and 8 moments $M_n$ as a
function of $Q^2$.  All the moments appear qualitatively similar,
rising to within about 10\% of their asymptotic values by $Q^2=1$
GeV$^2$.  Also evident is the fact that the lower moments reach
their asymptotic values earlier than the higher moments.  This is
qualitatively consistent with the expectation from the operator
product expansion discussed in \cite{DGP}, where it was argued
that the effective expansion parameter in the twist expansion
$\sim n/Q^2$, so that for higher moments, $n$, the higher twist
terms survive to larger values of $Q^2$.

\begin{figure}[!h,t]    
\begin{center}
\leavevmode
\epsfig{file = 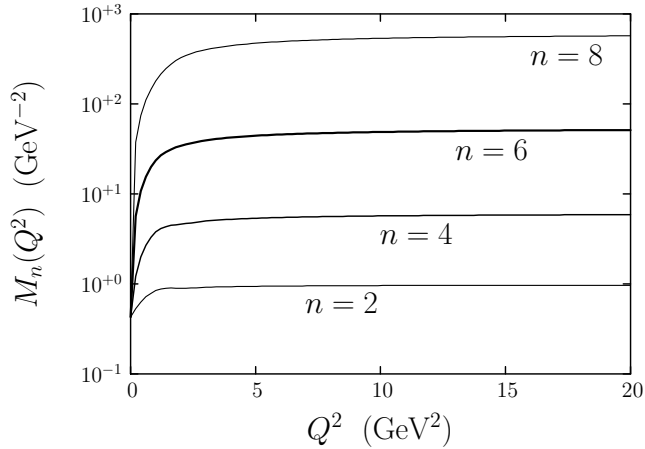, height=16cm}
\end{center}
\vspace{-10cm}
\caption{Some moments $M_n$ as a function of $Q^2$.}
\label{figmom1dist}
\end{figure}

Unfortunately, these moments do not have such useful interpretations here
as they do in real deep inelastic scattering. For example,
the analog of the Gross-Llewellyn Smith sum rule
is not applicable here because the scalar current  which
couples to our quark is not conserved.
Nonetheless, the moments in Fig. ~\ref{figmom1dist} do serve to
demonstrate that scaling is a natural consequence of our model, and
illustrate the relative onset of scaling for different moments.

\section{Onset of Scaling and Bloom-Gilman Duality}

After studying the scaling behavior of the structure functions in
our model at high $Q^2$ and the moments over a range of
four-momentum transfers, we now study the structure functions at
low $Q^2$ where not only in the large $N_C$ limit but also in
nature resonances are visibly dominant over a wide range in the
scaling variable. Here, we consider a target where only one quark
carries all the charge of the system, so there is no forced
breakdown of duality at $Q^2 = 0$   of the type noted earlier for
the neutron. Still, one cannot expect that the perturbative QCD
result will describe even averaged hadronic observables well at
very low $Q^2$: these are after all strong interactions!

If local duality holds, one might expect the resonance ``spikes" to
oscillate around the scaling curve and to average to it, once $Q^2$ is
large enough. (We remind the reader that while scaling in
deep-inelastic electron scattering from the nucleon is known from
experiment to set in by $Q^2 \sim 2$~GeV$^2$, the target considered
here corresponds to an infinitely heavy ``meson'' composed of scalar
quarks interacting with a scalar current, so one should not expect
numerically realistic results, only qualitative ones.)  Figure~3 shows
the onset of scaling for the structure function ${\cal{S}}_{cq}$ as a
function of $u$, as $Q^2$ varies from 0.5~GeV$^2$ to 2~GeV$^2$. As in
Fig.~1, for each of the resonances (excluding the elastic peak) the
energy $\delta$-function has been smoothed out using the Breit-Wigner
method with a width $\Gamma=100$~MeV.  With increasing $Q^2$, each of
the resonances moves out towards higher $u$, as dictated by
kinematics. At $Q^2 = 0$, the elastic peak is the only allowed state
and contributes about 44\% of the asymptotic value of $M_2$.  It
remains rather prominent for $Q^2$ = 0.5~GeV$^2$, though most of $M_2$
is by this point built up of excited states, and it becomes negligible
for $Q^2 \geq$ 2.0 GeV$^2$.  Remarkably, the curves at lower $Q^2$ do
tend to oscillate (at least qualitatively) around the scaling curve,
as is observed in proton data. Note that these curves are at fixed
$Q^2$, but sweep over all $\nu$. In a typical low energy experiment,
$\nu $ will also be limited; in such circumstances these curves still
apply, but they get cut off at the minimum value of $u$ that is
kinematically allowed. For another perspective on these curves, note
that $\vert \vec q \vert^2 = Q^2+\nu^2$ so for fixed $Q^2$, as $\nu$
is increased so that more and more highly excited states are created,
the struck quark is being hit harder and harder.

    In contrast, the structure function ${\cal{S}}_{Bj}$ when plotted
as a function of the scaled Bjorken variable $u_{Bj}$ shows very poor
duality between its low- and high-$Q^2$ behaviors, as seen in Fig.~4.
One of the reasons for this failure is that $x_{Bj}$ and
${\cal{S}}_{Bj}$ know nothing about the constituent quark mass, while
low energy free quark scattering certainly does, so the corresponding
pQCD cross section calculated neglecting the quark mass is simply
wrong at low energy.

\begin{figure}[!h,t]    
\begin{center}
\leavevmode
\epsfig{file = 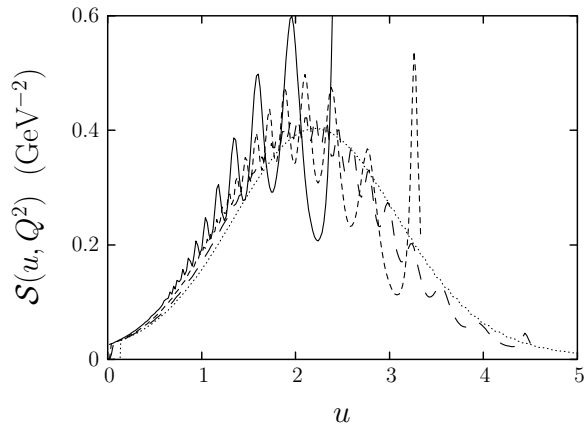, height=16cm}
\end{center}
\vspace{-10cm}
\caption{Onset of scaling for the structure function ${\cal S}_{cq}$
as a function of $u$ for $Q^2=0.5$ (solid), $Q^2=1$ (short-dashed), 2
(long-dashed) and 5~GeV$^2$ (dotted).  Although off-scale, the elastic
peak at $Q^2=0.5$ ~GeV$^2$ accounts for about 22\% of the area under
the scaling curve.  }
\label{figlocaldistbw}
\end{figure}

\begin{figure}[!h,t]    
\begin{center}
\leavevmode
\epsfig{file = 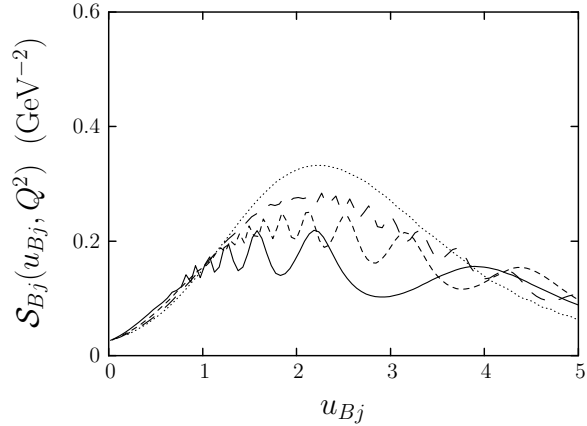, height=16cm}
\end{center}
\vspace{-10cm}
\caption{Onset of scaling for the structure function ${\cal S}_{Bj}$
as a function of $u_{Bj}$ for for $Q^2=0.5$ (solid), $Q^2=1$ (short-dashed),
2 (long-dashed) and 5~GeV$^2$ (dotted).}
\label{figlocalbjbw}
\end{figure}

\section{ Duality in Semileptonic Decays of Heavy Quarks}

     We have seen that low-energy (Bloom-Gilman) duality is displayed
by our model in terms of the appropriate low-energy variable $u$ and
described some of the physics behind this duality (completeness of
the bound state wave functions to expand a plane wave and an
approximate closure based on the required expansion states being
in a narrow band of $\nu$ relative to those that are kinematically
allowed).  To obtain a deeper understanding of the physics behind
low energy duality, it is instructive to compare and contrast
duality in electron scattering with that in heavy quark decays. We
will begin by carefully examining duality in heavy-light systems,
where it is exact in the heavy quark limit even down to zero
recoil, and where the mechanisms behind this exact duality are
very clear.

   Duality in heavy quark systems is easily understood intuitively.
Consider a $Q^* \bar q$ system where $m_Q^* >> \Lambda_{QCD}$, and imagine
that
$Q^*$ can decay to $Q$ by emitting a scalar particle $\phi$ of mass $\mu$:
$Q^*
\rightarrow Q+\phi$. (Note that in this case it is the heavy quark that
interacts with the current and not the light quark as in our model!) At the
free quark level, the decay of
$Q^*$ at rest will produce the $\phi$ with a single sharp kinetic energy
$T_{free}$ and  corresponding $Q$ recoil velocity $\vec v$. (We use the
standard variables
$T_{free}$ and $\vec v$, but others, like the $\phi$ recoil momentum, could
be chosen.) In reality, since the heavy quarks are bound into mesons, $\phi$
will (in the narrow resonance approximation)   emerge from the decay at
rest of  the initial meson's  ground state $(Q^* \bar q)_0$    with any of
the sharp kinetic energies allowed by the processes
$(Q^*
\bar q)_0 \rightarrow (Q \bar q)_n + \phi$ as determined by the strong
interaction spectra of these two mesonic systems. Since in the heavy quark
limit $m_{(Q^* \bar q)_n}-m_{(Q \bar q)_n} \simeq m_{Q^*}-m_Q$, $m_{(Q^*
\bar q)_n} \simeq m_{Q^*}$, and $m_{(Q
\bar q)_n} \simeq m_{Q}$,  the hadronic spectral lines are guaranteed to
cluster around $T_{free}$, and to coincide with it exactly as $m_Q
\rightarrow \infty$. Moreover, since
$m_Q^*,m_Q >>
\Lambda_{QCD}$, one can show using an analog of the operator product
expansion \cite{inclorig} that the strong interactions can be neglected in
calculating the total decay rate ({\it {\it i.e.}}, the heavy quarks  $Q^*$
and
$Q$ are so heavy  that the decay proceeds  as
though it  were free.) Thus the sum of the strengths of the spectral lines
clustering around $T_{free}$ is the free quark strength: there is perfect
low energy duality as $m_Q^*, m_Q
\rightarrow \infty$.

What is now especially interesting is to unravel
this duality to understand how the required ``conspiracy'' of spectral line
strengths   arises physically. Because the heavy quark is so massive, if it
would as a free particle recoil  with a velocity $\vec v$, then  this
velocity would be changed only negligibly by the strong interaction since in
the heavy quark limit  it carries off a negligible kinetic energy, but a
momentum much larger than $\Lambda_{QCD}$. In the rest frame of the
recoiling meson, this configuration requires that the two constituents have
a {\it relative} momentum $\vec q$ which grows with $\vec v$.  {\it Thus the
strong interaction dynamics is identical to that of our model in which the
relative momentum $\vec q$ is supplied by the scattered electron.} Moreover,
in this case, with duality exact at all energies, we can reconstruct
exactly how it arises. What one sees is remarkably simple
\cite{BjSumRule,IWonBj}. At low $\vec v$ corresponding to low
$\vec q$, only the ground state process $(Q^*
\bar q)_0 \rightarrow (Q \bar q)_0 + \phi$ occurs. Since the masses and
matrix elements for the transitions $(Q^*
\bar q)_0 \rightarrow (Q \bar q)_0 + \phi$ and $Q^*
  \rightarrow Q + \phi$ are identical (the elastic form factor goes
identically to unity as $\vec q \rightarrow 0$), the hadronic and quark
spectral lines and strengths are also identical and duality is valid at
$\vert \vec q
\vert^2=0$! Next consider duality at a different kinematic point (which one
might reach by choosing a smaller $\phi$ mass) where
$\vec v$ and therefore
$\vec q$ have increased. The elastic form factor will fall, so its
spectral line (which is still found at exactly the new value of  $T_{free}$
in the heavy quark limit) will carry less strength. However, once $\vec q $
differs from zero, excited states $(Q \bar q)_n$ can be created, and indeed
are created with a strength that exactly compensates for the loss of elastic
rate. These excited state   spectral lines   also coincide with
$T_{free}$  and duality is once again exact.
Indeed, no matter how large $|\vec q|^2$ becomes, all of the excited states
produce spectral lines at $T_{free}$ with strengths that sum to that of the
free quark spectral line.

Heavy quark theory also allows
one to go beyond the heavy quark limit to the case of quarks of finite
mass. In this case one of course finds that duality-violation occurs, but
that it is formally suppressed  by two powers of
$\Lambda_{QCD}/m_Q$ \cite{inclorig,DualityControversy}, with the
spectral lines now clustered about
$T_{free}$ but not coinciding with it.  A remarkable feature of this
duality violation is that the spectral line strengths   differ from
those of the heavy quark limit  in ways that tend to compensate for the
duality-violating phase space effects from the spread of spectral lines
around $T_{free}$. An additional source of duality-violation is
that some of the high mass resonances that are required for exact duality
are kinematically forbidden since for finite heavy quark masses
$m_{Q^*}-m_Q$ is finite.

    From this discussion it is   clear that the strong interaction
dynamics of heavy-light decays is the same as that of scattering a probe
off of the
$Q$ of a $Q \bar q$ system \cite{XpQCD}: what is relevant   is
that the system must in each case respond to a  relative momentum kick
$\vec q$.
Needless to say, one must still carefully organize the kinematics to expose
duality: in a decay to a
fixed mass $\phi$ only a single magnitude   $\vert \vec q
\vert^2 $ is produced at the
quark level, while in electron scattering a large range of $\vert \vec q
\vert^2 $ and $\nu$ is produced by a given electron beam.

    Given these connections, it is relevant to note that in addition to the
obvious conceptual relevance of heavy-light systems,   model studies
indicate that in  these systems heavy quark behavior continues to hold
qualitatively even for $m_Q \sim m$. These models are, as one might
expect, similar to ours which displays the same clustering of spectral
lines, the same tendency for excited state spectral lines to compensate for
the fall with $\vert \vec q
\vert^2 $ of lighter states, and the same sources of duality
violation such as kinematically forbidden states and mismatches between the
mass of the recoiling hadrons and the struck quark. We
believe that these elements of the dynamics are clearly in operation and
that we have understood through our model  that the qualitative
applicability of duality for real systems should indeed extend  all of the
way down to zero recoil as seen in Nature.

\section{Summary and Outlook}

We have presented a simple, quantum-mechanical model in which we were
able to qualitatively reproduce the features of Bloom-Gilman duality.
The model assumptions we made are the most basic ones possible: we
assumed relativistic, confined, valence scalar quarks and treated the
hadrons in the infinitely narrow resonance approximation.  To further simplify
the situation, we did not consider a three quark ``nucleon'' target, but a
target made up by an infinitely heavy antiquark and a light quark. The
present work does not attempt to quantitatively describe any data, but to
give qualitative insight into the physics of duality.

Our work complements previous work on duality, where the experimental
data were analyzed in terms of the operator product expansion (OPE)
\cite{JI,DGP}. There, it was observed that at moderate $Q^2$,
the higher twist corrections to the lower moments of the structure
function are small. The higher twist corrections arise due to initial
and final state interactions of the quarks and gluons.  Hence, the
average value of the structure function at moderate $Q^2$ is not very
different from its value in the scaling region.  While
true, this statement is merely a rephrasing in   the language of the
operator product expansion of the experimentally observed fact that the
resonance curve averages to the scaling curve. However,
the operator product expansion does not explain why a certain
correction is small or why there are cancellations: the expansion
coefficients which determine this behavior are not predicted. The numerical
confirmation of these coefficients will eventually  come from a numerical
solution of QCD on the lattice, but an {\it understanding} of the physical
mechanism that leads to the small values of the expansion coefficient will
almost certainly only be found in the framework of a model like ours.

For example, one clear lesson from our study of duality is that the
commonly made sharp distinction between the ``resonance region'',
corresponding to an invariant mass $W < 2 $ GeV for scattering from a
proton, and the deep inelastic region, where $W > 2 $ GeV, is
completely artificial.

    Finally, we remind the reader that our model, with all the charge
on a single quark, with scalar currents, and with no spin degrees of
freedom, leaves much to be done in model-building.  The next step is
to use more realistic currents. While making the calculations more
complicated, coupling to the conserved quark current will allow one to
study the $Q^2$-evolution of the Gross-Llewellyn Smith and momentum
sum rules. To use a spin-${1\over 2}$ target will also be a useful
step forward, but it may require foregoing the great advantages of the
analytic solutions of the Klein-Gordon equation. As we have
emphasized, the local duality seen here {\it cannot} be expected for
more complicated targets and processes, and pursuing this issue is
also clearly very important \cite{CloseIsgur}. Here we have taken a
first small step which nevertheless has been enough to strongly
suggest that for these more realistic models and more general
processes there will be a generalization of local averaging --- a
theoretically well-defined procedure for integrating over regions of
$x_{cq}$ --- which will also display low energy duality. If so, we
will not only have understood quark-hadron duality. We will also have
opened the door to extending studies of a variety of structure
functions into previously unreachable kinematic regimes.

\acknowledgments

We thank C.~Carlson, F.~Close, R.~Ent, J. Goity, C.~Keppel, R.~Lebed,
I.~Niculescu and S.~Liuti for stimulating discussions, and
N.N.~Nikolaev and S.~Simula for pointing out several references.  This
work was supported in part by DOE contract DE-AC05-84ER40150 under
which the Southeastern Universities Research Association (SURA)
operates the Thomas Jefferson National Accelerator Facility (Jefferson
Lab), and by the Australian Research Council.

\end{document}